\documentclass[aps,prstab,twocolumn,showpacs,superscriptaddress,groupedaddress]{revtex4}  

\usepackage{dcolumn}   
\usepackage{bm}        
\usepackage{amssymb}   
\usepackage{color}

\newcommand{\secref}[1]{Section~\ref{#1}}
\newcommand{\figref}[1]{Fig.~\ref{#1}}
\newcommand{\tblref}[1]{Table~\ref{#1}}
\newcommand{\eqnref}[1]{Eq.~(\ref{#1})}

 \newcommand{\sang}{\ensuremath{\sin \theta}}
 \newcommand{\cang}{\ensuremath{\cos \theta}}
 \newcommand{\bz}{\ensuremath{\beta_z}}
 \newcommand{\bx}{\ensuremath{\beta_x}}
 \newcommand{\ax}{\ensuremath{\alpha_x}}

\usepackage{color}

\usepackage{amsmath}
\usepackage{verbatim} 
\usepackage{color}
\usepackage[section] {placeins}
\usepackage{graphicx}  

\usepackage[caption=false,font=footnotesize]{subfig}

\begin{document}

\title{Aberration Corrected Emittance Exchange}
\author{E.~A.~Nanni} \affiliation{Massachusetts Institute of Technology, Cambridge, MA 02139, USA}
\author{W.~S.~Graves} \affiliation{Massachusetts Institute of Technology, Cambridge, MA 02139, USA}

%
%
 \noaffiliation%
\vskip 0.25cm

\date{\today}

\begin{abstract}
Full exploitation of emittance exchange (EEX) requires aberration-free performance of a complex imaging system including active radio-frequency (RF) elements which can add temporal distortions. We investigate the performance of an EEX line where the exchange occurs between two dimensions with normalized emittances which differ by multiple orders of magnitude. The transverse emittance is exchanged into the longitudinal dimension using a double dog-leg emittance exchange setup with a five cell RF deflector cavity. Aberration correction is performed on the four most dominant aberrations. These include temporal aberrations that are corrected with higher order magnetic optical elements located where longitudinal and transverse emittance are coupled. We demonstrate aberration-free performance of an EEX line with emittances differing by four orders of magnitude, \textit{i.e.} an initial transverse emittance of 1~pm-rad is exchanged with a longitudinal emittance of 10~nm-rad.
\end{abstract}

\pacs{}
\maketitle
\section{Introduction}

Emittance exchange (EEX) \cite{cornacchia2002transverse,sun2010tunable,carlsten2011using,emma2006transverse} is the process by which the emittance (phase-space) of an electron bunch in two dimensions is fully coupled and exchanged. Potential applications include bunch compressors \cite{carlsten2011using}, phase space manipulations for x-ray free electron lasers (FEL) \cite{emma2006transverse} and  current profile shaping for plasma or dielectric wakefield acceleration \cite{piot2011generation,jiang2012formation,Rosenzweig1988,thompson2008}. Additionally, this innovative geometry led to the proposed exchange of a transversely modulated electron beam for the use of x-ray generation via coherent inverse Compton scattering \cite{graves2012intense,gravescompact} or by seeding a FEL \cite{jiang2011emittance}. On the mm scale this transverse modulation can be imparted by a slotted mask \cite{sun2010tunable,jiang2011emittance} and can be used for a THz radiation source \cite{piot2011generation}. On the nanometer scale, the production of coherent x-ray beams has been proposed by transferring a spatial electron modulation from a nanocathode \cite{graves2012intense}  into a temporal modulation using emittance exchange.  

Inherently, most applications desire the EEX line to exchange a relatively large emittance with a small one. The exchange of transversely patterned beams into a longitudinal nanobunching is an excellent example of the extreme mismatch in emittance between the initial transverse emittance of one nanobunch compared to its longitudinal emittance. For the modulation period required for coherent x-ray generation this ratio approaches 10$^4$. Each nanobunch originates within a small fraction of the total beam, therefore it has a small fraction of the total emittance, yet that small emittance must be fully exchanged with (nearly) the total emittance in the longitudinal plane.  Residual coupling with the original dimension, resulting in incomplete exchange and subsequently emittance growth, limits the performance of the EEX line. Detailed first order analysis has led to EEX configurations that can fully exchange the emittance in two dimensions when considering first order analytical treatments \cite{emma2006transverse,piot2011generation}. However, there has been no detailed analysis of the aberrations present in the system which result in emittance growth and set the ratio of emittances which can be exchanged. Additionally, aberration corrected imaging has not been implemented to improve the performance of EEX.  Aberration correction in low energy electron optics has been driven by the quest to image individual atoms using electron microscopes, where sub-angstrom resolution \cite{batson2002sub} has been achieved. In electron microscopes, optics are typically limited to systems with no temporal dependence due to the low currents and lack of time varying electromagnetic fields. In EEX, the optics include temporal aberrations which we find can be corrected with  higher order magnetic optical elements located where longitudinal and transverse emittance are coupled. 

In this paper we present the aberration corrected geometry for an EEX beamline and demonstrate exchange of transverse and longitudinal emittances which differ by four orders of magnitude. This extreme difference between transverse and longitudinal emittance is representative of a single nanobunch in a transversely modulated beam. We only consider aberrations from externally applied fields such as quadrupole magnets and radio-frequency (RF) deflecting cavities.  The beam's space charge field may also result in phase space distortions and emittance growth, but these effects depend on the particular beam being considered and require numerical studies beyond the scope of the present work.  \secref{sec:EEXdynamics} details the general first order theory for an EEX beamline and the matching conditions required at its input to produce an electron bunch that is uncorrelated in time and energy at its output. A comparison between the analytical and numerical performance for an EEX beamline is discussed in \secref{sec:firstorderperformance}. \secref{sec:HOC} explains the dominant sources of aberrations in the EEX beamline and the placement of aberration-correcting magnetic elements. Numerical simulations are shown in \secref{sec:ebeamsim} for the performance of the EEX line using an electron bunch with and without the presence of aberration-correcting elements. Concluding remarks are given in \secref{sec:conclusions}.

\begin{figure*}[t]
  \includegraphics[width=0.9\textwidth]{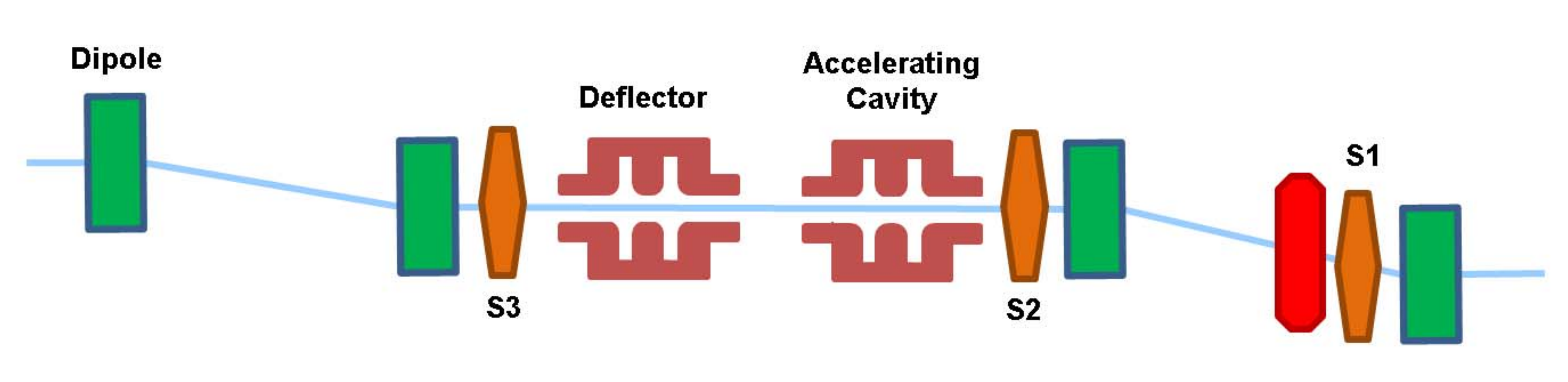}
  \caption {Schematic of the EEX line showing the primary components of the system. It consists of two doglegs separated by a short straight section containing deflecting and accelerating RF cavities.  The electron beam enters from the left with the reference electron trajectory shown with a blue line. Green squares represent bending magnets, orange hexagons represent sextupoles and the red octagon represents an octupole.  The total length is approximately 2 m.}
  \label{fig:layout}
\end{figure*}

\section{Emittance exchange beam dynamics}
\label{sec:EEXdynamics}
In the present discussion, the EEX line consists of two doglegs separated by a transverse deflecting cavity and a longitudinally accelerating cavity. We examine the case where the EEX line is used to transform a small portion of the transverse beam distribution (beamlet) into an upright ellipse (\textit{i.e.} with no time-energy correlation) in the z-dimension at the exit of the EEX. At the input of the EEX line the beamlet's transverse emittance in the x-dimension, $\epsilon^\text{in}_x$, is much less than its longitudinal emittance $\epsilon^\text{in}_z$. Each particle passing through the setup can be described by its 6D phase-space coordinates
			\begin{equation}
			X^\text{T}=\left( \begin{matrix}  x &  x' & y & y' & z & {\Delta p}/{p} \end{matrix} \right)
		\end{equation}
where all terms are defined with respect to a reference particle, $x$ and $y$ are the transverse coordinates of the particle, $x'$ and $y'$ are the transverse angular divergence, $z=\beta c t$ is the longitudinal position, $p=\beta \gamma m_ec$ is the momentum, $c$ is the speed of light, $\beta=v/c$, $v$ is the electron velocity, $m_e$ is the electron mass and $\gamma=1/\sqrt{(1-v^2/c^2)}$ is the Lorentz factor. We define a first order transfer matrix $R$ for the EEX line given by the transfer matrices of the subcomponents ${{X}_\text{out}}=R{{X}_\text{in}}={{R}_\text{d}}{{R}_\text{a}}{{R}_\text{k}}{{R}_\text{d}}{{X}_\text{in}}$ \cite{carey1998third}. The primary components of the EEX line consist of two translating sections or doglegs ($R_\text{d}$) composed of opposite polarity dipoles and a drift space, separated by a transverse deflecting cavity ($R_\text{k}$) and an accelerating cavity ($R_\text{a}$) which are both operated at zero crossing RF phase. At zero crossing the reference particle receives no net deflection or acceleration. A schematic with a layout of the principal components of the EEX line is shown in \figref{fig:layout}. The doglegs and the deflecting cavity both provide a strong transverse-longitudinal coupling. As will be shown below, with the appropriate dispersion and deflector strength we achieve complete exchange of the longitudinal and transverse emittance. The only function of the accelerating cavity is to remove the temporal energy chirp induced by the deflector. The discussion of higher order effects due to aberrations and their mitigation with the use of sextupoles and an octupole (shown in \figref{fig:layout}) are presented in \secref{sec:HOC}.

The EEX occurs between the x-z dimensions only; ideally the y-dimension is uncoupled.  In a practical case the coupling to the y-dimension must be minimized. For clarity, the analytical expression for the transfer matrix and our discussion will be limited to the x-z dimensions so that rows 3 and 4 and columns 3 and 4 will not be shown for the general $6 \times 6$ transfer matrix. Considering only the x-z dimensions that are exchanged, matrix transport describing the EEX line is given by the $4\times 4$ matrix $R = \bigl ( \begin{smallmatrix} A & B  \\ C & D \end{smallmatrix}\bigr )$ consisting of $2 \times 2$ submatrices $A$, $B$, $C$ and $D$.  In most beamlines there is no coupling between the transverse and longitudinal phase space dimensions so that $B=C= \bigl ( \begin{smallmatrix} 0 & 0  \\ 0 & 0 \end{smallmatrix}\bigr )$.  However the EEX line is designed to exactly exchange phase space dimensions so that now the components of $B$ and $C$ are nonzero but the components of submatrices $A=D= \bigl ( \begin{smallmatrix} 0 & 0  \\ 0 & 0 \end{smallmatrix}\bigr )$.

For the ideal case the transfer matrix of the dogleg is 
		\begin{equation}
		{{R}_\text{d}}=\left( \begin{matrix}
   1 & L_\text{d} & 0 & \eta   \\
   0 & 1 & 0 & 0  \\
   0 & \eta  & 1 & \zeta   \\
   0 & 0 & 0 & 1  \\
\end{matrix} \right)
\end{equation}
where $L_\text{d}$ is the drift length, $\eta $ is the horizontal dispersion and $\zeta$ is the longitudinal dispersion. The drift length is
\begin{equation} \label{eq:Ld}
L_\text{d} = \frac{L_D }{\cos^3 \theta} + \frac{2 L_M}{\cang} + L_\text{drift},
\end{equation}
the horizontal dispersion is
\begin{equation} \label{eq:eta}
\eta = \frac{L_D \sang}{\cos^2 \theta} + \frac{2 L_M}{\sang} \left(\frac{1}{\cang}-1\right),
\end{equation}
and the longitudinal dispersion is
\begin{equation} \label{eq:zeta} 
\zeta = \frac{L_D \sin^2 \theta}{\cos^3 \theta} + 2 L_M \left( \frac{1}{\cang} - \frac{\theta}{\sang} \right) + \frac{L_\text{d}}{\gamma^2 \beta^2}
\end{equation}
where $\theta$ is the bend angle, $L_D$ is the distance between dipoles in each dogleg, $L_M$ is the length of the dipoles and $L_\text{drift}$ is the distance between the dipole and the RF cavities.  These terms closely follow the derivation in \cite{carlsten2011using} with the important difference of the added last term in \eqnref{eq:zeta} that measures momentum dependent path length.  This term becomes important for electrons that are only moderately relativistic.  It describes the path length difference $\Delta z = 1/(\gamma^2\beta^2) (\Delta p/p)$ due to velocity differences and is critical to account for when attempting to use an EEX line at energies of a few MeV to tens of MeV with small emittance beams.  

For the deflector 
		\begin{equation}
		{{R}_\text{k}}=\left( \begin{matrix}
   1 & D & \frac{kD}{2} & 0  \\
   0 & 1 & k & 0  \\
   0 & 0 & 1 & 0  \\
   k & \frac{kD}{2} & \frac{k^2D}{6} & 1  \\
\end{matrix} \right)
\end{equation}
where $k$ is related to the change in transverse momentum of the particle and $D$ is the length of the deflector. Emittance exchange requires the RF deflector cavity strength to scale inversely with horizontal dispersion $k = -1/\eta$ where $k = \omega_\text{RF} V_{\bot} / c E$, $V_{\bot}$ is the integrated deflecting voltage, $\omega_\text{RF}$ is RF frequency and $E$ is beam energy.

In order to correct for $R^\text{k}_{65}$ we add an accelerating cavity operated at zero crossing \cite{shin2012modeling} with the transfer matrix
		\begin{equation}
		{{R}_\text{a}}=\left( \begin{matrix}
   1 & d & 0 & 0  \\
   0 & 1 & 0 & 0  \\
   0 & 0 & 1 & 0  \\
   0 & 0 & -\chi & 1  \\
\end{matrix} \right)
\end{equation}
where $\chi=\frac{2\pi}{\lambda}\frac{|e|V_{\parallel}}{pc}$, $V_{\parallel}$ is the integrated accelerating voltage, $\lambda$ is the RF wavelength and $d$ is the length of the accelerating cavity. If we set the strength of the accelerating cavity such that $\chi=-\frac{k^2D}{6}$ then the combined cavities have a transfer matrix
\begin{equation}
		{{R}_\text{a}{R}_\text{k}}=\left( \begin{matrix}
   1 & D+d & \frac{kD}{2}+kd & 0  \\
   0 & 1 & k & 0  \\
   0 & 0 & 1 & 0  \\
   k & \frac{kD}{2} & 0 & 1  \\
\end{matrix} \right).
\end{equation}

For the ideal case, the deflector strength $k={-1}/{\eta }$ which results in complete emittance exchange. The transfer matrix of the full EEX line is 
	\begin{equation}
	R=\left( \begin{matrix}
   0 & 0 & -\frac{L+d}{\eta } & \eta -\frac{(L+d)\zeta }{\eta }  \\
   0 & 0 & -\frac{1}{\eta } & -\frac{\zeta }{\eta }  \\
   -\frac{\zeta }{\eta } & \eta -\frac{\zeta L}{\eta } & 0 & 0  \\
   -\frac{1}{\eta } & -\frac{L}{\eta } & 0 & 0  \\
\end{matrix} \right)
\end{equation}
where $L=L_\text{d}+D/2$.

Transport through the EEX line for the entire electron bunch can be represented in matrix notation as
\begin{equation} \label{eq:rsigrt}
\Sigma_{z} = R \Sigma_{x} R^T.
\end{equation}
Here the incoming transverse-beam phase-space ellipse, $\Sigma_{x}$, at the first EEX dipole is 
\begin{equation}
\Sigma_{x} = \begin{pmatrix} \bx & -\ax \\
-\ax & \gamma_{x} \end{pmatrix}
\end{equation}
where \ax, \bx~and $\gamma_{x}=\frac{1+(\ax)^2}{\bx}$ are the Twiss parameters. $\Sigma_{x}$ is transformed into the final upright longitudinal beam ellipse containing no time-energy correlation
\begin{equation}
\Sigma_{z} = \begin{pmatrix} \bz & 0 \\
0 & \frac{1}{\bz} \end{pmatrix}.
\end{equation}

To find the incoming \ax~and \bx~required to match \bz,~\eqnref{eq:rsigrt} is inverted 
\begin{equation} \label{eq:rinv}
\Sigma_{x} = R^{-1} \Sigma_{z} (R^T)^{-1}.
\end{equation}
Using det($B$) = det($C$) = 1 and solving for the elements of interest
\begin{eqnarray}
\bx & = & \bz R_{62}^2 + \frac{1}{\bz} R_{52}^2 \nonumber \\
  & \approx & \frac{1}{\bz} R_{52}^2 \label{eq:bx} \\
\ax & = & \bz R_{61} R_{62} + \frac{1}{\bz} R_{51} R_{52} \nonumber \\
  & \approx & \frac{1}{\bz} R_{51} R_{52} \label{eq:ax}
\end{eqnarray}
where the dropped terms are negligible.  From \eqnref{eq:bx} it is seen that the incoming beta function scales with $R_{52}^2$ which in turn is dominated by the velocity dependent longitudinal dispersion term of \eqnref{eq:zeta} for electron beams of a few tens of MeV energy.  This dispersion scales linearly with beamline length thus it is important to minimize the total length of the EEX line in order to control the incoming beta function.

Under the stated condition $\epsilon^\text{in}_x\ll\epsilon^\text{in}_z$, it is critical for the values in the lower right quadrant of $R$ to be nearly zero. An approximate requirement on the first order transport matrix of the EEX line can be determined by considering the output particle distribution in the longitudinal dimension which is given by

\begin{equation}
\left( \begin{matrix} z  \\ \Delta p/p \end{matrix} \right)= C\left( \begin{matrix} x  \\ x' \end{matrix}\right) + D\left( \begin{matrix} z  \\ \Delta p/p \end{matrix}\right).
\label{eq:rmatrixreqspec}
\end{equation}

In order for $\epsilon^\text{out}_z \approx  \epsilon^\text{in}_x$, the contribution to the longitudinal dimensions must satisfy $D\bigl ( \begin{smallmatrix} z  \\ \Delta p/p \end{smallmatrix}\bigr ) \ll C\bigl ( \begin{smallmatrix} x  \\ x' \end{smallmatrix}\bigr )$.  Without considering a specific particle distribution, but rather the ratio of emittances that must be exchanged we find the approximate requirement
\begin{equation}
D_{m,n}/C_{m,i}<\epsilon^\text{in}_x/\epsilon^\text{in}_z
\label{eq:rmatrixreq}
\end{equation}
This important result is the guiding design principle for optimizing the transport through the EEX.

\section {First Order Performance}
\label{sec:firstorderperformance}

Given the ideal performance for the EEX line and its subcomponents presented in \secref{sec:EEXdynamics}, it is important to asses a realistic layout for the EEX line. We must also incorporate the impact of additional couplings due to weak thick lens effects that were disregarded in \secref{sec:EEXdynamics}. These could have a significant impact when considering emittances which differ by orders of magnitude. In particular, the expected result $A=D= \bigl ( \begin{smallmatrix} 0 & 0  \\ 0 & 0 \end{smallmatrix}\bigr )$ is not strictly true, leading to emittance growth for the electron bunch primarily in the dimension of interest, $\epsilon^{\text{out}}_z$.

\begin{table}[t]
\caption{Emittance Exchange Line Parameters \label{tab:eex} }
\begin{center}
\begin{tabular}{l c c c c}\hline\hline\
Parameter & Symbol & Value &  Unit \\
\hline
Operational Energy  &  & 22.5  & MeV  \\
Dipole-dipole drift & $L_D$ & 1  & m  \\
Dipole length & $L_M$ & 6   & cm  \\
Dogleg-Deflector drift & $L_\text{drift}$ & 5 & cm \\
Bend angle   & $\theta$ &  10    & degrees  \\
Deflector strength & $k_{rf}$ &  5.27 &  $\mathrm{m}^{-1}$  \\
Deflector voltage  & $V_{\bot}$ &   0.6083    & MeV \\
Central drift length  &  & 26.44   &  cm \\
Dogleg dispersion & $\eta$ & 18.97  & cm \\
Total EEX length & $2L$  &  2.5 & m  \\
\hline \hline
\end{tabular}
\end{center}
\vspace{-3mm}
\end{table}

\begin{table}[t]
\caption{RF Deflector Parameters \label{tab:deflector}}
\begin{center}
\begin{tabular}{l c c }\hline\hline\
Parameter &  Value &  Unit \\
\hline
RF frequency & 9300 & MHz \\
Number of Cells &   5  & -  \\
R$_\text{T}$ Single cell & 0.35  & M$\Omega$  \\
R$_\text{T}$ Total & 1.3  & M$\Omega$  \\
\text{Q$_\text{L}$} &  5000   & -  \\
\text{V$_\perp$} &  0.7 & MeV\\
\text{P}$_\text{in}$ &   200    & kW  \\
\text{E}$_\text{max}$ &32 & MV/m\\
\text{H}$_\text{max}$ &112 & kA/m\\
\hline \hline
\end{tabular}
\end{center}
\vspace{-3mm}
\end{table}

The nominal design parameters of the EEX line were selected for generation of coherent current modulation at a few tens of MeV keeping in mind the overall size, requirements for RF power and input electron bunch parameters. The full parameters of the EEX line elements are summarized in \tblref{tab:eex}.  These design choices and linear performance will be discussed in the following subsections. 

The first order performance for each subcomponent is determined by solving the linear system of equations ${{X}_n^\text{out}}={{R}}{{X}_n^\text{in}}$ for the transport matrix ${{R}}$, where $n$ is the particle index for the electron bunch, ${X}_n^\text{in}$ is the input particle phase-space distribution, and ${X}_n^\text{out}$ is the output distribution generated by tracking ensembles of particles through the beamline elements with the simulation program \textsc{parmela}. This overdetermined linear system, where $n\gg 6\times 6$, is conveniently solved by Matlab functions that minimize the Frobenius norm residual.


\subsection{Dogleg} 

It was shown in \eqnref{eq:bx} that the incoming beta function scales with $R_{52}^2$ and that this dispersion scales linearly with beamline length. The two doglegs are the dominant contributors to size in the system. From \eqnref{eq:eta} the dispersion can be adjusted by varying the bend angle $\theta$ and dipole-dipole drift length $L_D$.  Keeping the drift short and the bend angle modest, the desired dispersion is achieved with a bend angle of $\theta =  10$ degrees and drift length $L_D = 1$ m. The analytical and numerical transfer matrix for the dogleg is shown in \tblref{tab:rmatricies}.

\begin{table*}[p]
\caption{First Order Transfer Matrix \label{tab:rmatricies} }
\begin{center}
\begin{tabular}{l c c c }\hline\hline\
Element & Symbol& Analytical & Numerical \\
\hline
Dogleg & ${R}_\text{d}$   &$\left( \begin{matrix}
   1 & 1.2188 & 0 & 0.1897   \\
   0 & 1 & 0 & 0  \\
   0 & 0.1897  & 1 & 0.0334  \\
   0 & 0 & 0 & 1  \\
\end{matrix} \right)$ & $\left( \begin{matrix}
   1  & 1.2028  & 0 & 0.1897   \\
   0  & 1 & 0 & 0  \\
   0 & 0.1897   & 1 & 0.0329    \\
   0  & 0  & 0 & 1   \\
\end{matrix} \right)$  \\

Deflector Cavity & ${R}_\text{k}$  &$\left( \begin{matrix}
   1 & 0.108 & -0.2846 & 0  \\
   0 & 1 & -5.27 & 0  \\
   0 & 0 & 1 & 0  \\
   -5.27 & -0.284 & 0.370 & 1  \\
\end{matrix} \right)$ & $\left( \begin{matrix}
1	& 0.108		& -0.2847		& 0 \\
1.55\times10^{-4}	& 	1		& -5.27		& 0\\
-1.39\times10^{-4}		& 0		& 1		& 0\\
-5.27		& -0.286		& 0.336		& 1\\
\end{matrix} \right)$  \\

Accelerating Cavity & ${R}_\text{a}$   &$\left( \begin{matrix}
   \ \ 1 \ \ & 0.0564 & 0 & 0  \\
   0 & 1 & 0 & 0  \\
   0 & 0 & 1 & 0  \\
   0 & 0 & -0.3705 & 1  \\
\end{matrix} \right)$ & $\left( \begin{matrix}
   1 & 0.0564  & 0 & 0  \\
		-6.0\times10^{-4} & 1 & 0 & 0 \\
   0 & 0 & 1 & 1.13\times10^{-4}  \\   
	-3.21\times10^{-3} & 	1.17\times10^{-3} & -0.329 & 1  \\
\end{matrix} \right)$  \\

EEX Beamline & ${R}$  &$\left( \begin{matrix}
   0 & 0 & -7.27 & -0.0530  \\
   0 & 0 & -5.27 & -0.176  \\
   -0.176 & -0.043 & 0 & 0  \\
   -5.27 & -6.97 & 0 & 0  \\
\end{matrix} \right)$ & $\left( \begin{matrix}
   -1.13\times10^{-3} & -7.58\times10^{-4} & -6.92 & -0.034 \\
   -5.80\times10^{-3} &-1.05\times10^{-3} & -5.27 & -0.155  \\
   -0.174 & -0.029 & -3.72\times10^{-6} & -1.54\times10^{-5}  \\
   -5.33 & -6.63 & 1.72\times10^{-3} & -3.69\times10^{-4}  \\
\end{matrix} \right)$  \\

\hline \hline
\end{tabular}
\end{center}
\vspace{-3mm}
\end{table*}

\begin{figure*}[p]
   \centering
   \includegraphics*[width=0.85\textwidth]{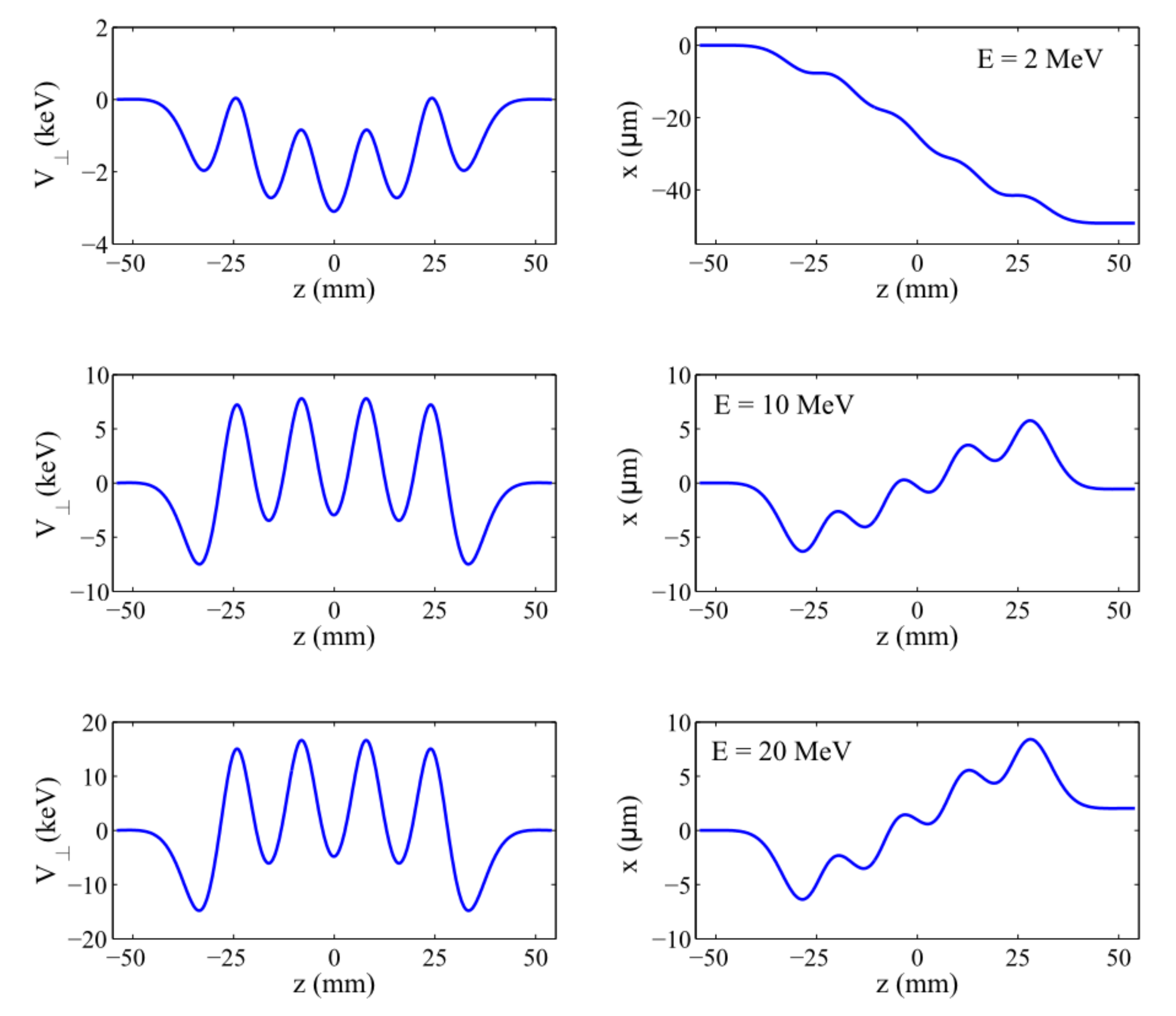}
   \caption{Transverse momentum and the transverse offset as a function of distance in the deflector cavity for the reference particle at zero-crossing. The deflector cavity strength at 2, 10 and 20~MeV corresponds to a $V^{\text{max}}_\perp$ = 0.07, 0.35 and 0.7~MeV, respectively. The final offset of the beam at the exit of the cavity is optimized for 10~MeV. If the beam energy is increased/decreased the final transverse offset shifts to positive/negative values.}
   \label{fig:deflector}
\end{figure*}

\subsection{RF Structures}
A 5 cell deflector cavity with a standing wave mode to minimize the required RF power was designed for use in the EEX line. Due to the low-energy electron beam, unique design features were included in the deflector to minimize the transverse offset at zero crossing. Additionally, the cavity was optimized to operate over the 2-25~MeV range accessible by a compact linac. The TM$_{110}$ mode is excited by two couplers located in the center cell. The transverse shunt impedance \cite{alesini2006rf}, defined as 
\begin{equation}
{{R}_{T}}=\frac{{{\left| \int_{0}^{{D}}{\left( c{{B}_{x}}+{{E}_{y}} \right){{e}^{{j{{\omega }_\text{RF}}z}/{c}\;}}dz} \right|}^{2}}}{2P}
\end{equation} 
where $B_x$ and $E_y$ are the transverse magnetic and electric fields in the cavity and $P$ is the power coupled into the cavity, determines the the maximum deflecting voltage for a given input power. The design parameters for the deflector are shown in \tblref{tab:deflector}. For EEX the deflector cavity operates at zero crossing imparting opposite kicks to the leading and trailing particles. Under a thin lens approximation the deflector cavity does not produce a transverse offset for the reference particle at zero crossing. However, due to the finite length of the cavity the reference particle would traverse the deflector cavity with a non-zero mean transverse momentum \cite{floettmann2014beam}, resulting in a transverse offset. In order to mitigate this each of the cavity end cells provides half the kick of a full cell. This effectively results in a zero mean transverse momentum for a single optimized energy. In \figref{fig:deflector} the transverse particle momentum and offset are shown for the reference particle at three energies (2, 10, 20~MeV) covering the full range of interest. The final transverse momentum for all three cases is zero due to proper phasing of the reference particle. However, the mean transverse offset is non-zero for the 2 and 20~MeV case. This is demonstrated by the non-zero transverse offset for the particle at the exit of the cavity. For the EEX parameters listed in \tblref{tab:eex} with a dispersion of $\eta = 18.97$~cm and a 22.5~MeV beam, the RF power required for the deflector is 150~kW.

The accelerating cavity is a standard 3 cell TM$_{01}$ standing-wave structure operated at the zero crossing. The RF frequency  is 9300~MHz and the full structure length is 5.64~cm. The drive power for the cavity is just 265~W and the shunt impedance is 208~M$\Omega$/m. The analytical and numerical transfer matrices for both the deflector and the accelerating cavity are shown in \tblref{tab:rmatricies}.

\subsection{Full EEX Line}

Combining all the elements of the EEX line we find the analytical and numerical performance in \tblref{tab:rmatricies}. The transfer matrix contains some residual self coupling for both the transverse and longitudinal dimension. The performance of the EEX line was optimized to minimize the residual self coupling in the longitudinal dimension, $D \Rightarrow \bigl ( \begin{smallmatrix} 0 & 0  \\ 0 & 0 \end{smallmatrix}\bigr )$, as this dimension will contain the small final emittance. We previously stated an approximate requirement in \eqnref{eq:rmatrixreq} that determines the maximum emittance ratio which can be exchanged and we find that this requirement is generally met for an exchange ratio of four orders of magnitude. As we will see in \secref{sec:ebeamsim}, full numerical simulations for the electron bunch will demonstrate this large exchange in emittance and validate the strict requirement given in \eqnref{eq:rmatrixreqspec} which also considers the electron bunch distribution for exchange without emittance growth.

\section{Aberrations and their Corrections}
\label{sec:HOC}

While the linear performance of the EEX line demonstrates an excellent ability to exchange emittances which differ by moderate ratios, higher-order couplings eventually lead to emittance growth that limits the achievable ratio. The first order matrix formalism applied in \secref{sec:EEXdynamics} to analyze the performance of the EEX line can be expanded to include 2$^\text{nd}$ and 3$^\text{rd}$ order transfer matrices in order to analyze the coupling of aberrations into the output emittance and determine the placement of corrective higher-order optics. Particle trajectories in 6D phase space can be described as
\begin{equation}
	{{{X}}_\text{out}}={{R}}{{{X}}_\text{in}}+{{{T}}}{X}_\text{in}{X}_\text{in}+{{{{M}}}}{X}_\text{in}{X}_\text{in}{X}_\text{in}+...
\end{equation}
where ${{T}}$ and ${{M}}$ are the 2$^\text{nd}$ and 3$^\text{rd}$ order matrices respectively.  These higher order terms may be treated as an error in the linear transport with terms given by 
\begin{equation}
	 {{\Delta X}_{i}}=\sum\limits_{j,k}{{{T}_{ijk}}{{X}_{j}}{{X}_{k}}}+\sum\limits_{j,k,l}{{{M}_{ijkl}}{{X}_{j}}{{X}_{k}}{{X}_{l}}}.
\end{equation}

Given the large ratio of emittances and the significant coupling between the x-z dimension throughout the EEX line, it is difficult to observe small aberrations that only produce measurable errors when the full emittance exchange occurs after the final dipole. To increase the visibility of these higher-order couplings, the performance of the EEX line was analyzed by producing artificial electron bunches with a distribution of particles that has finite width in only one dimension, and zero width in the others. This allows us to observe the initial appearance of higher-order couplings due to aberrations, how these errors propagate through the EEX optics and their impact on the final emittance. After identifying the source of the aberrations and placing correcting optics, we then return to using the nominal electron bunch at the entrance of the EEX line.

The dominant aberrations in the EEX line are related to the temporal field variation of the deflector cavity. Due to the low energy and finite length of the electron bunch and the alternating fields in the deflector cavity, quadratic and cubic errors appear in the transfer matrix. Due to the coupling of the transverse and longitudinal emittance in the EEX line it is possible to correct these errors without additional time-dependent electron optics.  This is achieved with the use of sextupoles and octupoles placed where the quadratic or cubic dependence of the static magnetic element in the transverse dimension can compensate for the corresponding aberration. A similar approach with the placement of sextupoles in highly dispersive regions was used in \cite{england2005sextupole} to correct for energy spread induced aberrations in dispersionless translating sections. However, in that case the aberrations were the result of static magnetic optics as opposed to dynamic electromagnetic structures. Sextupoles and octupoles are a natural choice for correcting aberrations because their lowest order transport, 2$^\text{nd}$ and 3$^\text{rd}$ order respectively, contain only geometric terms allowing coupling between longitudinal and transverse dimensions via horizontal dispersion and avoiding additional 2$^\text{nd}$ and 3$^\text{rd}$ order chromatic aberrations, respectively.

The strongest of the aberrations that appears in the deflector cavity is the quadratic dependence on energy due to the length of the bunch, ${\Delta p}/{p}=T_{655}^\text{k}z_\text{k}^2$. Here we use superscripts on transfer matrix elements (\textit{e.g.} k, d,...) and subscripts on particle coordinates to identify the relevant component or location in the system.   This quadratic aberration results in emittance growth for the output beam $\epsilon^\text{out}_z$ because it is coupled through the dogleg into the temporal dimension, $z_\text{out}^\text{error}=R_{56}^\text{d}T_{655}^\text{k}z_\text{k}^2$. We can correct this error with a sextupole (S1) placed between the two bending magnets of the second dogleg. The sextupole will create a quadratic coupling between the transverse extent of the beam and the divergence, $x'=T_{211}^{S1}x_\text{S1}^2$. This quadratic dependence can create a correcting term in the temporal dimension, $z_\text{out}^\text{correction}=R_{52}^\text{dipole}T_{211}^\text{S1}x_\text{S1}^2$. However, we must show that the dimension $x$ at the sextupole  is correlated with the temporal dimension $z$ at the deflector. Following the transfer matrix from the deflector we find the dominant correlation to be  $z_\text{out}^\text{correction}=R_{52}^\text{dipole}T_{211}^\text{S1}(L_\text{prop}R_{25}^\text{k})^2z_\text{k}^2$ where $L_\text{prop}$ is the total distance between the end of the deflector and the sextupole location. Therefore, to remove the aberration we set 
\begin{equation}
R_{56}^\text{d}T_{655}^\text{k}=-R_{52}^\text{dipole}T_{211}^\text{S1}(L_\text{prop}R_{25}^\text{k})^2.
\end{equation}
We place the sextupole immediately prior to the final dipole which minimizes the magnetic field strength needed to correct for the aberration. 

Through a nearly identical argument we are also able to correct the cubic non-linearity in the deflector cavity, ${\Delta p}/{p}=M_{6555}^\text{k}z_\text{k}^3$, with the placement of an octupole prior to the final dipole and setting the strength of the octupole to equal
\begin{equation}
R_{56}^\text{d}M_{6555}^\text{k}=-R_{52}^\text{dipole}M_{2111}^\text{O}(L_\text{prop}R_{25}^\text{k})^3.
\end{equation}

The second quadratic temporal aberration of the deflector cavity that we will address is the $T_{255}^\text{k}$. This aberration results in an error at the output due to its coupling to the z-dimension through the second dogleg $z_\text{out}^\text{error}=R_{52}^\text{d}T_{255}^\text{k}z_\text{k}^2$. We can correct for this aberration with a sextupole (S2) placed after the accelerating cavity, $z_\text{out}^\text{correction}=R_{52}^\text{d}T_{211}^sx_\text{S2}^2$. This placement avoids 3$^\text{rd}$ order chromatic effects because the accelerating cavity removes the z-dependent chirp on the electron bunch.  At this location the transverse extent of the beam, $x_\text{S2}=(R_{15}^\text{k}+L_\text{prop2}R_{25}^\text{k})z_{k}$,  is linearly correlated to the longitudinal dimension at the entrance of the deflector. $L_\text{prop2}$ is the distance from the end of the deflector cavity to the sextupole S2. This gives the matching condition 
\begin{equation}
T_{255}^\text{k}=-T_{211}^\text{S2}(R_{15}^\text{k}+L_\text{prop2}R_{25}^\text{k})^2
\end{equation}
for removing the aberration.

\begin{table}[t]
\caption{Aberration Correcting Elements \label{tab:HOplacement} }
\begin{center}
\begin{tabular}{l c c}\hline\hline\
Optical Element & Position (cm) &  Strength  \\
\hline
Sextupole (S1) & 0  & -675 (Gauss/cm$^2$) \\
Sextupole (S2) & 4  & 356  (Gauss/cm$^2$)\\
Sextupole (S3) & 0  & -498 (Gauss/cm$^2$)\\
Octupole  & 0  & 58.6  (Gauss/cm$^3$)\\

\hline \hline
\end{tabular}
\end{center}
\vspace{-3mm}
\end{table}

\begin{table}[t]
\caption{General Electron Beam Parameters \label{tab:ebeam} }
\begin{center}
\begin{tabular}{l c c c}\hline\hline\
Parameter & Value &  Unit \\
\hline
Normalized emittance & $10^{-8}$  & m-rad  \\
RMS $\mathrm{\Delta p/p}$  &  $10^{-5}$    & --  \\
Energy  &  22.5 & MeV  \\
RMS bunch length & 69.4 &  fs  \\
Relativistic $\beta$ & 0.9998 &  \\
Relativistic $\gamma$ & 45.03 & \\

\hline \hline
\end{tabular}
\end{center}
\vspace{-3mm}
\end{table}

The third sextupole in the system accounts for aberrations related to the finite transverse size of the electron bunch that is due to the initial transverse emittance and the finite energy spread of the beam. The transverse beam size after the first dogleg and at the entrance of the third sextupole is $x_\text{S3}=x_\text{in}+R_{12}^\text{d}x_\text{in}'+R_{16}^\text{d}{\Delta p_\text{in}}/{p}$. At this point we note that several downstream elements result in quadratic correlations between $x_\text{S3}$ and the transverse divergence. These elements are the deflector and both sextupoles placed to correct temporal aberrations. The impact of these elements depends strongly on the transverse size of the beam at each individual element's location and the strength of each element. As was shown previously, the effects of each element are cumulative, noting that the quadratic correlations produced by these elements alter the z-dimension at the output. We can refer to this aberration as $T_{211}^\text{eff}$ and correct it with the placement of a sextupole at the entrance of the deflector and setting the strength to $T_{211}^\text{eff}=-T_{211}^\text{S3}$.

The physical placement of the sextupoles and octupoles is shown in \figref{fig:layout} and the location and strength of each element for our design is given in \tblref{tab:HOplacement}, where the position of each element is given as the distance from its end to the start of the subsequent downstream element. The length of the sextupoles and the octupole are 1~cm each.

\begin{table}[t]
\caption{Emittance and Twiss Parameters \label{tab:ebeam_modeling} }
\begin{center}
\begin{tabular}{l c c c}
\hline\hline\
Parameter & x &  y &z \\

\hline
\\
\multicolumn{4}{c}{EEX Input} \\
\hline

Emittance (nm-rad) & 0.001  & 10 & 10  \\
$\alpha$ & 49.7  & 0 & 0  \\
$\beta$ (m) & 8.3  & 3.0 & 2.0  \\
RMS ($\mu$m) & 0.45  & 25.7 & 21.1  \\

\hline
\\
\multicolumn{4}{c}{EEX Output} \\
\hline
Emittance (nm-rad) & 10.0  & 10.0 & 0.0055  \\
$\alpha$ & -73.2  & -0.27 & -0.04 \\
$\beta$ (m) & 95.7  & 0.91 & 5.1$\times10^{-4} $ \\
RMS ($\mu$m) & 145.9  & 14.26 & 0.00788  \\

\hline
\\
\multicolumn{4}{c}{EEX Output with Aberrations Corrected} \\
\hline
Emittance (nm-rad) & 13.5  & 10.0 & 0.001  \\
$\alpha$ & -54.1  & -0.27 & -0.04  \\
$\beta$ (m) & 70.8  & 0.91 & 1.0$\times10^{-4}$ \\
RMS ($\mu$m) & 145.9  & 14.26 & 0.00148  \\

\hline \hline
\end{tabular}
\end{center}
\vspace{-3mm}
\end{table}

\section{Aberration corrected performance}
\label{sec:ebeamsim}
\begin{figure}[t]
  \includegraphics[trim = 0cm 0mm 0cm 0mm, width=0.47\textwidth]{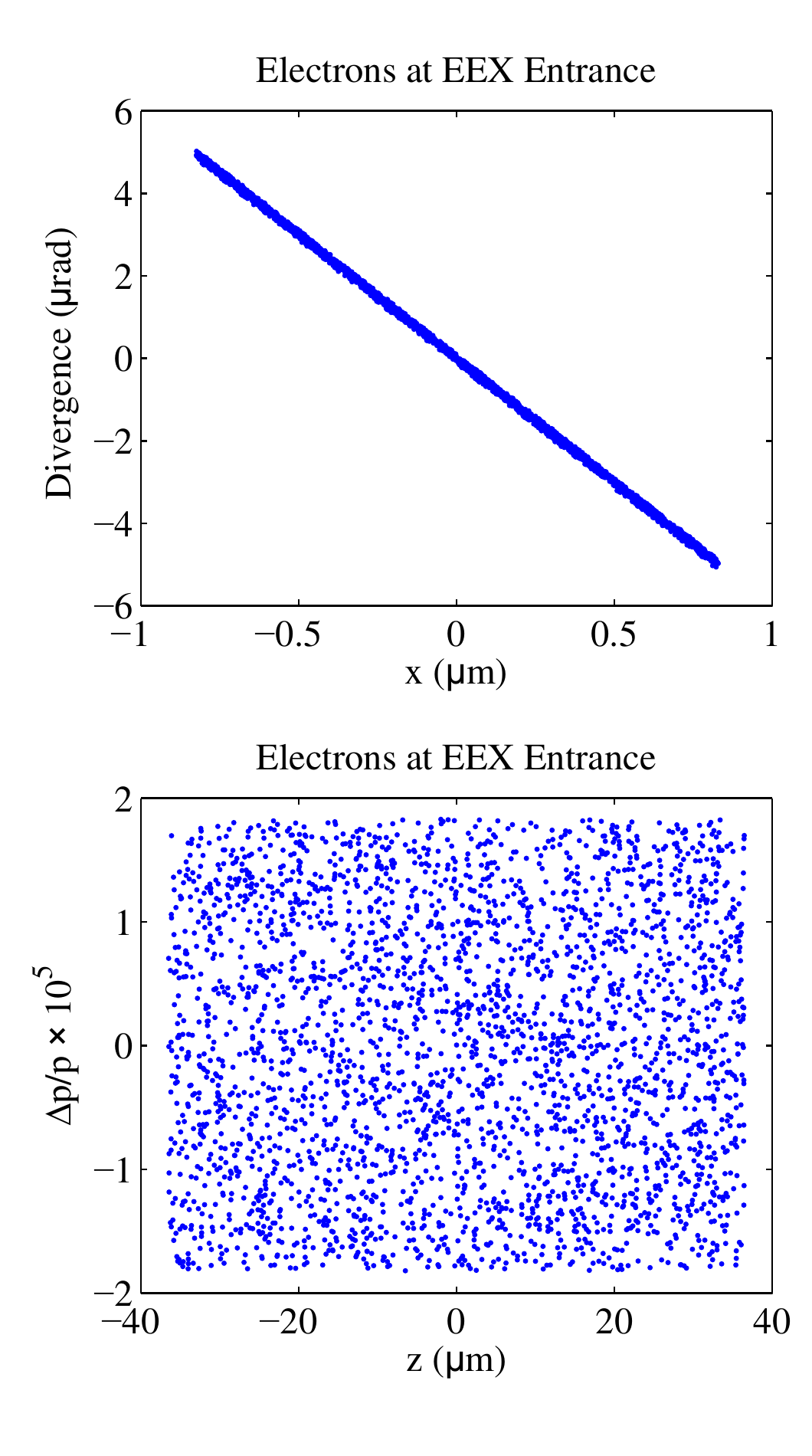}
  \caption {The phase space of the input electron bunch in the (top) transverse and (bottom) longitudinal dimension. The input parameters were selected to produce a beam with no energy chirp at the exit.}
  \label{fig:EEXinput}
\end{figure}

To test the EEX line we use an electron bunch with an initial transverse emittance of $\epsilon_x=1$~pm-rad and longitudinal emittance of $\epsilon_z=10$~nm-rad. The mean energy of the electron bunch is  22.5~MeV. The normalized transverse emittance in the y-dimension is $\epsilon_y=10$~nm-rad. These electron beam parameters are consistent with a nanobunch generated to produce coherent soft x-ray radiation \cite{graves2012intense,nanninanometer}.  The electron beam properties at the input of the EEX line are summarized in \tblref{tab:ebeam} and the input emittances and Twiss parameters are listed in \tblref{tab:ebeam_modeling}. We show the particle phase space at the input of the EEX line for the two dimensions of interest in \figref{fig:EEXinput}. A randomly populated square distribution in phase-space is chosen to ease the visualization of aberrations in the electron bunch.

\begin{figure}[t]
  \includegraphics[width=0.49\textwidth]{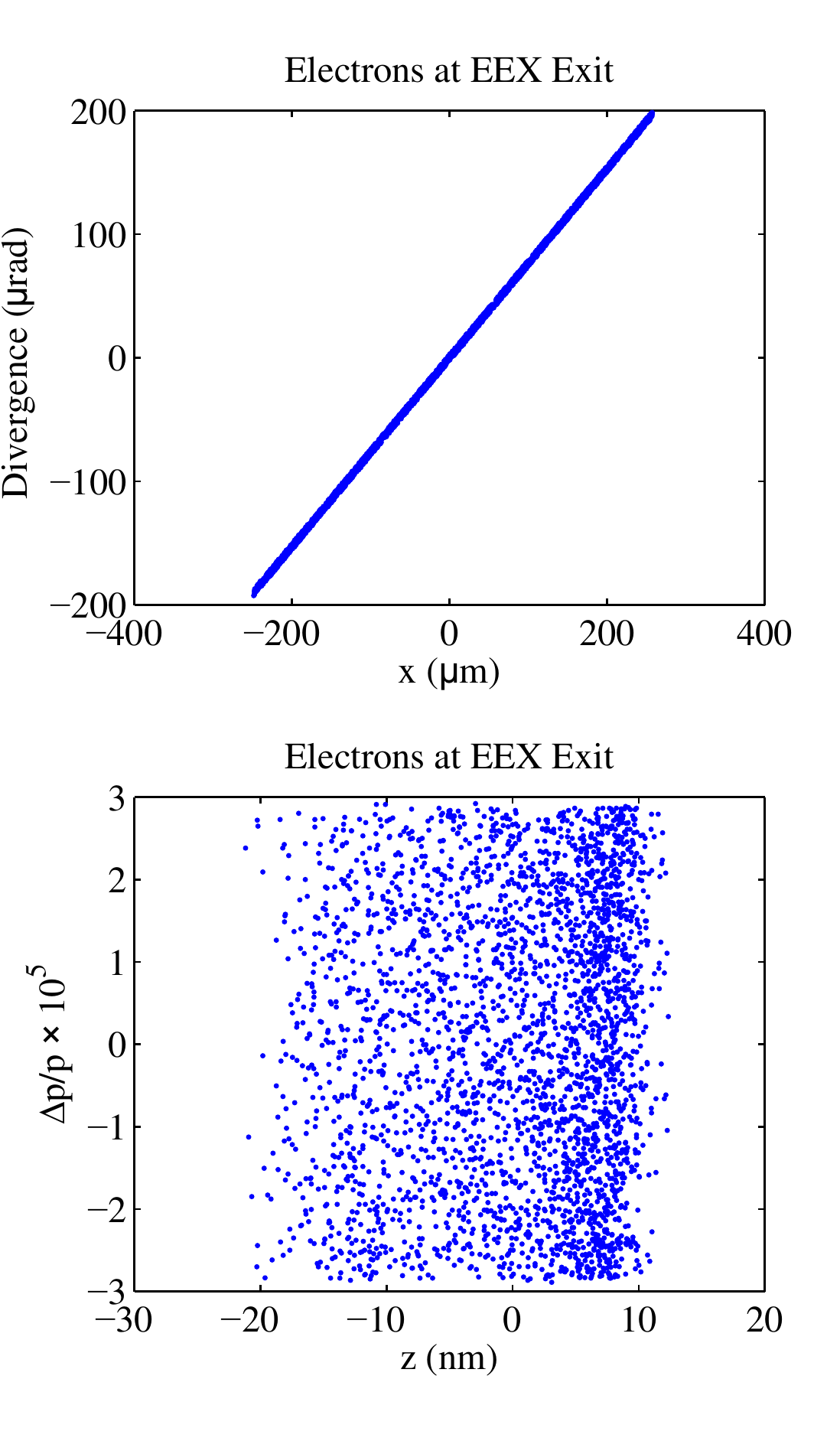}
  \caption {The phase space of the output electron bunch in the (top) transverse and (bottom) longitudinal dimension when the aberration correcting optics are turned off. A large emittance growth is observed, primarily due to errors that are coupled into the longitudinal dimension of the electron bunch.}
  \label{fig:EEXaberration}
\end{figure}

\begin{figure}[t]
  \includegraphics[trim = 0cm 0mm 0mm 0mm, width=0.50\textwidth]{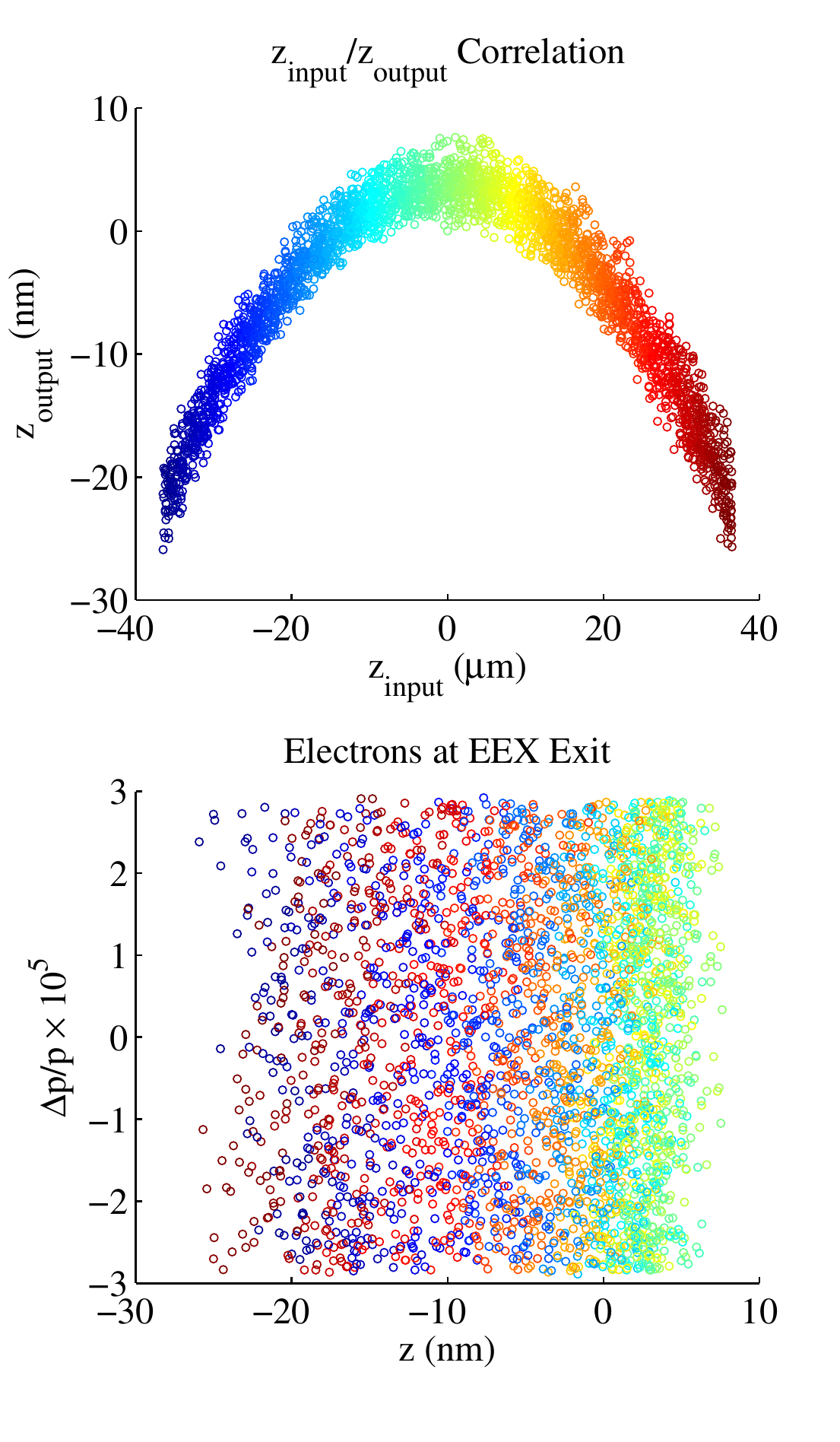}
  \caption {(top) Correlation between the longitudinal position at the entrance to the transverse deflector cavity and at the exit of the EEX line. Particles are color coded to the input position. (bottom) Individual particles in the longitudinal phase space. The quadratic longitudinal coupling dominates the emittance growth of the electron bunch.}
  \label{fig:EEX_dominantabb}
\end{figure}

We begin the analysis of the performance for the full EEX line without correcting for aberrations by zeroing the 3 sextupoles and the octupole. \figref{fig:EEXaberration} shows the particle phase space at the output of the EEX line for the two dimensions of interest. The emittance and Twiss parameters for the output electron bunch are shown in \tblref{tab:ebeam_modeling}. We note that without aberration correction the EEX line is unable to preserve the initial emittance $\epsilon^\text{in}_x$, resulting in a final emittance $\epsilon^\text{out}_z$ that is a factor of 5.5 larger. The emittance growth is dominated by the $T^\text{k}_{655}$ and the $T^\text{k}_{255}$ aberrations. This quadratic aberration is shown in \figref{fig:EEX_dominantabb} where the particle position at the output of the EEX line with respect to the longitudinal position at the entrance of the deflector is shown in color-coded scatter plots. 


\begin{figure}[t]
  \includegraphics[trim = 0cm 0mm 0mm 0mm, width=0.49\textwidth]{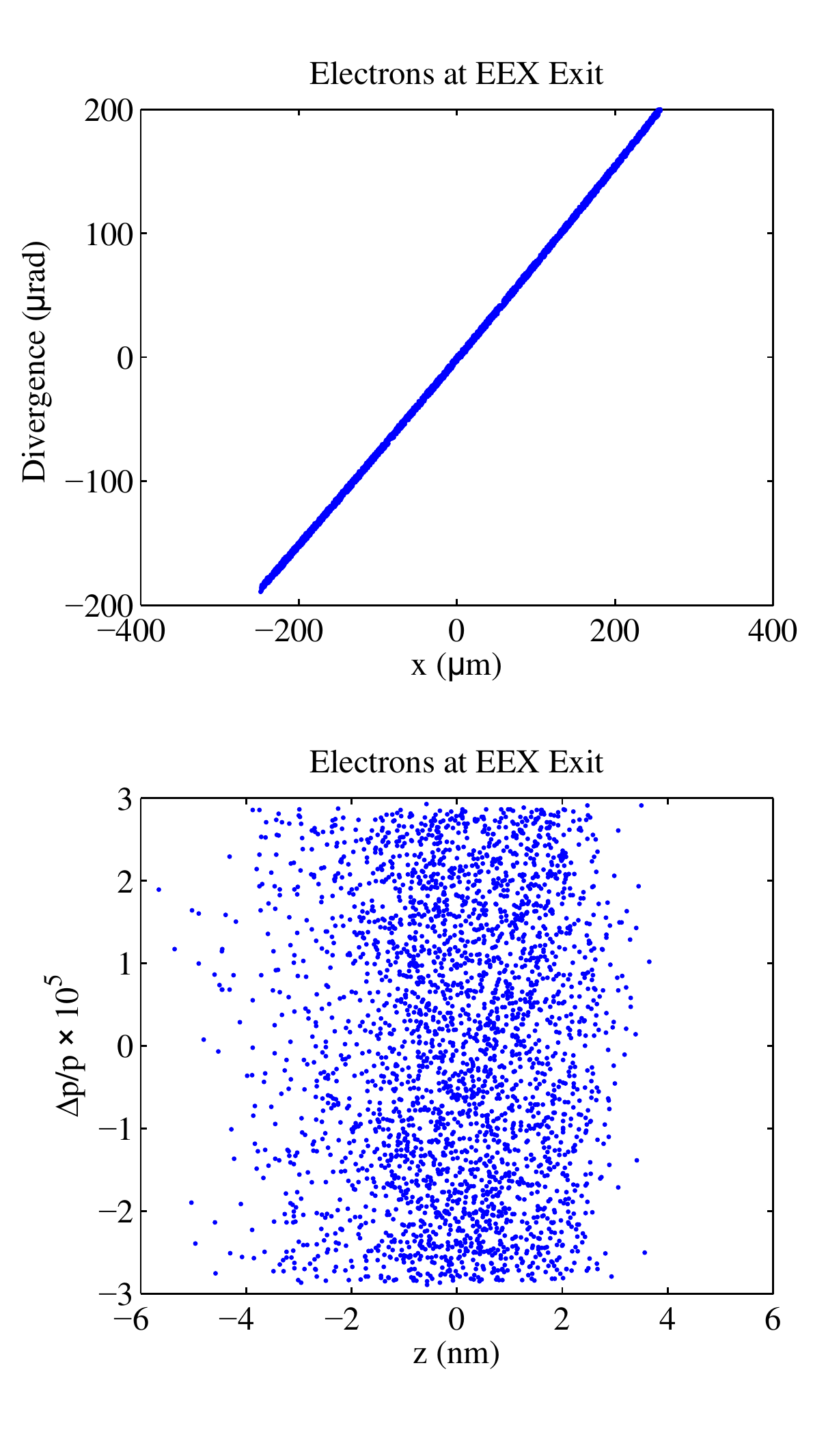}
  \caption {The phase space of the output electron bunch in the (top) transverse and (bottom) longitudinal dimension.}
  \label{fig:EEXoptimal}
\end{figure}

We now include all the aberration correcting sextupoles and octupole in the EEX line with the particle phase space at the output for the two dimensions of interest shown in \figref{fig:EEXoptimal}. The emittance and Twiss parameters for the output electron bunch are shown in \tblref{tab:ebeam_modeling}. The EEX line performs exceptionally well when considering the final z emittance, which now shows zero emittance growth. The RMS length, $\sigma_z$, of the beam at the output is only 1.5~nm which is four orders of magnitude shorter than the input electron beamlet.

\section{Conclusions}
\label{sec:conclusions}
We have presented the design and simulation of a compact emittance exchange line capable of transforming emittance ratios as large as 10,000 between the transverse and longitudinal dimensions of an electron beam.  The exchange of such large ratios is a necessary step toward periodic nanobunching of an electron beam at low energy.  Such an approach will enable the efficient generation of coherent FEL-like radiation at much lower energies than traditional approaches, reducing the size and cost of future light sources while improving their output radiation properties.  

\clearpage



Successful emittance exchange requires optimization of the linear transport, especially the time-varying fields of the RF deflector cavity.  We have shown that aberrations due primarily to these time varying fields ultimately limit the ability to exchange arbitrarily large ratios of emittance, and that these aberrations may be mitigated through the use of higher order static sextupole and octupole magnets correctly positioned to take advantage of the coupling among multiple dimensions in the EEX.

In this analysis only the dominant aberrations were discussed and corrected with the use of static magnetic elements. However, given different objectives for the EEX line or different electron beam parameters other aberrations can play an equal or more significant role. In particular we note that doglegs can introduce quadratic aberrations in the longitudinal position as a function of energy and in $x$ as function of $x'$. Both of these aberrations can be corrected with the placement of sextupoles in the doglegs. Additionally, x-y coupling in higher-order optical elements may also present itself as a source of emittance growth and should be mitigated with the choice ${\beta}_y$ in these elements or with additional corrective optics prior to the EEX line. 

The end result is successful emittance exchange and thus the ability to generate nanoscale bunches with achievable electron beam parameters.

\clearpage

\section{Acknowledgements}
\label{sec:acknowledgements}

 The authors gratefully acknowledge Philippe Piot for many useful discussions. 
 
 This work was supported by NSF Grant No. DMR-1042342, DOE Grant No. DE-FG02-10ER46745 and DARPA Grant No. N66001-11-1-4192.

\bibliography{PhDbib}
\bibliographystyle{apsrev}

\end{document}